\documentclass[journal]{IEEEtran}
\usepackage{cite}

\ifCLASSINFOpdf
  \usepackage[pdftex]{graphicx}
\else
  \usepackage[dvips]{graphicx}
\fi
\usepackage{amsmath}
\usepackage{array}
\usepackage{multirow}
\usepackage{booktabs}
\usepackage{color,xcolor}
\usepackage{bm}

\begin{document}

\title{PGMAN: An Unsupervised Generative Multi-adversarial Network for Pan-sharpening}

\author{Huanyu~Zhou,
        Qingjie~Liu,~\IEEEmembership{Member,~IEEE,}
        Yunhong~Wang,~\IEEEmembership{Fellow,~IEEE}
\thanks{Authors are with the State Key Laboratory of Virtual Reality Technology and Systems, Beihang University and Hangzhou Innovation Institute, Beihang University. Email: \{zhysora, qingjie.liu, yhwang\}@buaa.edu.cn. Qingjie Liu is the corresponding author of this paper. This work was supported by NSFC Nos. 41871283 and 61601011.}
}

\markboth{Journal of Selected Topics in Applied Earth Observations and Remote Sensing,~Vol.~1, No.~1, Dec.~2021}%
{Shell \MakeLowercase{\textit{et al.}}: Bare Demo of IEEEtran.cls for IEEE Journals}

\maketitle

\begin{abstract}
Pan-sharpening aims at fusing a low-resolution (LR) multi-spectral (MS) image and a high-resolution (HR) panchromatic (PAN) image acquired by a satellite to generate an HR MS image. Many deep learning based methods have been developed in the past few years. However, since there are no intended HR MS images as references for learning, almost all of the existing methods down-sample the MS and PAN images and regard the original MS images as targets to form a supervised setting for training. These methods may perform well on the down-scaled images, however, they generalize poorly to the full-resolution images. To conquer this problem, we design an unsupervised framework that is able to learn directly from the full-resolution images without any preprocessing. The model is built based on a novel generative multi-adversarial network. We use a two-stream generator to extract the modality-specific features from the PAN and MS images, respectively, and develop a dual-discriminator to preserve the spectral and spatial information of the inputs when performing fusion. Furthermore, a novel loss function is introduced to facilitate training under the unsupervised setting. Experiments and comparisons with other state-of-the-art methods on GaoFen-2, QuickBird and WorldView-3 images demonstrate that the proposed method can obtain much better fusion results on the full-resolution images. Code is available\footnote{https://github.com/zhysora/PGMAN}.
\end{abstract}


\IEEEpeerreviewmaketitle

\section{Introduction}

\IEEEPARstart{D}ue to physical constraints \cite{physicalConstraint}, many satellites such as QuickBird, GaoFen-1, 2, WorldView I, II only offer a pair of modalities at the same time: multi-spectral (MS) images at a low spatial resolution and panchromatic (PAN) images at a high spatial resolution but a low spectral resolution. In many practical applications, it is desired to use high-resolution (HR) MS images. Pan-sharpening, which combines the strengths of an MS image with a PAN image to generate an HR MS image, provides a good solution to this.

Over the past few decades, researchers in the remote sensing community have developed various methods for pan-sharpening. These methods, to distinguish them from the recently proposed deep learning models we called them traditional pan-sharpening methods, can be mainly divided into three categories: 1) Component substitution (CS)-based methods, 2) Multi-resolution analysis (MRA)-based methods, and 3) Model-based methods. The CS methods convert MS images into a new space, in which one component is replaced with the spatial parts of the PAN, and then perform an inverse transformation to obtain the pan-sharpened images. Intensity-hue-saturation technique (IHS-based methods \cite{IHS}) , principal component analysis (PCA-based methods \cite{PCA1, PCA2}), and Gram-Schmidt (GS \cite{GS} method) are those widely adopted transformations. The MRA methods apply multi-resolution algorithms to extract the spatial information of PAN images and then inject them into MS images. Some representative methods include the modulation transfer function (MTF \cite{MTF1, MTF2}), and the smoothing filter-based intensity modulation (SFIM \cite{SFIM}). The model-based methods attempt to build interpretable mathematical models of the input PAN and MS and the ideal HR MS. They usually need to solve an optimization problem to parameterize the models. A typical method is band-dependent spatial detail (BDSD \cite{BDSD}) model. These traditional methods are widely used in practice, however, their ability to solve high nonlinear mapping is limited and thus often suffer from spatial or spectral distortions. 

Recently, deep learning techniques have made great successes in various computer vision tasks, from low-level image processing to high-level image understandings \cite{PercepLoss, RCAGAN, InstanceNorm}. Convolution neural networks (CNNs) have shown a powerful ability of modeling complex non-linear mappings and the superiority of solving image-enhancing problems such as single image super-resolution \cite{SR1, SR2}. Inspired by this, many deep learning models have been developed for pan-sharpening. PNN \cite{PNN} introduces SRCNN \cite{SRCNN} to pan-sharpening and designs a fusion network which is also a 3-layered CNN. PanNet \cite{PanNet} borrows the idea of skip-connection from ResNet \cite{ResNet} to build deeper networks and trains the model on the high-frequency domain to learn the residual between the up-sampled LR MS image and the desired HR MS image. DRPNN \cite{DRPNN} also learns from ResNet \cite{ResNet} and designs a deeper CNN with 11 layers. MSDCNN \cite{MSDCNN} tries to explore the multi-scale structures from images by using different sizes of filters and combining a shallow network and a deep network. TFNet \cite{TFNet} builds a two-stream fusion network and designs a variant of UNet \cite{Unet} to solve the problem. PSGAN \cite{PSGAN} improves the TFNet \cite{TFNet} by using generative adversarial training \cite{GAN}. 

These deep learning methods for pan-sharpening have achieved satisfactory performances. However, they cannot be optimized without supervised images and hence are hard to obtain optimal results on the full-resolution images. To be specific, the existing works require ideal HR MS images which do not exist to train networks. To optimize the networks, they down-sample PAN and MS images and take the original MS image as targets to form training samples. In the testing stage, evaluations are also conducted on the down-sampled images. However, for remote sensing images, this protocol may cause a gap between down-sampled images and the original ones.  Different from natural images, remote sensing images are usually in deeper bit depths and with distinct pixel distributions. These supervised methods may have good performances in the down-sampled image domain, however, they generalize poorly to the original full-scale images, which makes them lack practicality. 

To overcome this drawback, we propose an unsupervised generative multi-adversarial network, termed PGMAN. PGMAN focuses on unsupervised learning and is trained on the original data without down-sampling or any other pre-processing steps to make full use of original spatial and spectral information. Our method is inspired by CycleGAN~\cite{CycleGAN2017}. We use a two-stream generator to extract modality-specific features from PAN and MS images, respectively. Since we do not have target images to calculate losses, the only way for us to verify the quality of the generated images is the consistency property between the pan-sharpened images and the PAN and MS images, $i.e.$, the degraded versions of the HR MS images, both spectral and spatial degradations, should be as close as possible to the PAN and MS images. To realize this, we build two discriminators, one is to distinguish the down-sampled fusion results from the input MS images, and the other one is to distinguish the grayed fusion images from the input PAN. Furthermore, inspired by the non-reference metric $QNR$ \cite{QNR}, we introduce a novel loss function to boost the quality of the pan-sharpened images. Our major contributions can be summarized as follows: 
\begin{itemize}
\item We design an unsupervised generative multi-adversarial network for pan-sharpening, termed PGMAN, which can be trained on the full-resolution PAN and MS images without any pre-processing. It takes advantage of the rich spatial and spectral information of the original data and is consistent with the real application environment. 
\item For the purpose of being consistent with the original PAN and MS images, we transform the fusion result back to PAN and LR MS images and design a dual-discriminator architecture to preserve the spatial and spectral information.  
\item Inspired by the $QNR$ metric, We introduce a novel loss to optimize the network under the unsupervised learning framework without reference images. 
\item We conduct extensive experiments on GaoFen-2, QuickBird and WorldView-3 images to compare our proposed model with the state-of-the-art methods. Experimental results demonstrate that the proposed method can achieve the best results on the full-resolution images, which clearly show its practical value. 
\end{itemize} 
 
 The remainder of this paper is organized as follows. The related works and background knowledge are introduced in Section~\ref{section::Related Work}. The details of our proposed method are described in Section~\ref{section::Method}. Section~\ref{section::Experiments and Results} shows the experiments and the results. Finally, the conclusions are given in Section~\ref{section::Discussion}.

\section{Related Work} \label{section::Related Work}

Deep learning techniques have achieved great success in diverse computer vision tasks, inspiring us to design deep learning models for the pan-sharpening problem.  Observing that pan-sharpening and single image super-resolution share a similar spirit and motivated by \cite{SRCNN}, Masi et al. \cite{PNN} proposes a three-layered convolutional neural network (CNN) based pan-sharpening method. Following this work, increasing research efforts have been devoted to developing deep learning based pan-sharpening. For instance, Zhong et al. \cite{hybridPNN} present a CNN based hybrid pan-sharpening method. Recent studies \cite{deeper1, deeper2} have suggested that deeper networks will achieve better performance on vision tasks. The first attempt at applying the residual network is PanNet \cite{PanNet}. They adopt a similar idea to \cite{residualcnnforpansharp1} and \cite{residualcnnforpansharp2} but employ ResNet \cite{ResNet} to predict details of the image. In this way, both spatial and spectral information could be preserved well.

Generative adversarial networks (GANs), proposed by Goodfellow et al. \cite{GAN}, have achieved attractive performance in various image generation tasks. The main idea of GANs is to train a generator with a discriminator, adversarially. The generator learns to output realistic images to cheat the discriminator while the discriminator learns to distinguish the generated images from real ones. However, the difficulty of stable training of GANs remains a problem. DCGAN \cite{DCGAN} introduces CNN to GANs and drops the pooling layer, which improves the performance. LSGAN \cite{LSGAN} replaces the sigmoid cross entropy loss function with the least squares loss function to avoid vanishing gradients problem. WGAN \cite{WGAN} leverages the Wasserstein distance as the objective function and uses weight clipping to stabilize the training process. WGAN-GP \cite{WGAN-GP} penalizes norms of gradients of discriminators with respect to its input rather than uses weight clipping. SAGAN \cite{SAGAN} adds self-attention modules for long-range dependency modeling. To speed up convergence and ease the training process, we choose WGAN-GP as the basic GAN to build our model. 

Recently, researchers do not limit GANs in one-generator-one-discriminator architecture and try to design multiple generators and discriminators for dealing with difficult tasks. GMAN \cite{GMAN} extends GANs to multiple discriminators and endows them with two roles: formidable adversaries and forgiving teachers. One is a stronger discriminator while another is weaker. CycleGAN \cite{CycleGAN2017} designs two pairs of generators and discriminators and proposes a cycle consistency loss to reduce the space of possible mapping functions. MsCGAN \cite{MsCGAN} is a multi-scale adversarial network consisting of two generators and two discriminators handling different levels of visual features. SinGAN \cite{SinGAN} uses a pyramid of generators and discriminators to learn the multi-scale patch distribution in a single image. Considering the domain-specific knowledge of pan-sharpening, we design two discriminators to train against one generator for spectral and spatial preservation.

\begin{figure*}[t]
  \centering
  \includegraphics[width=\linewidth]{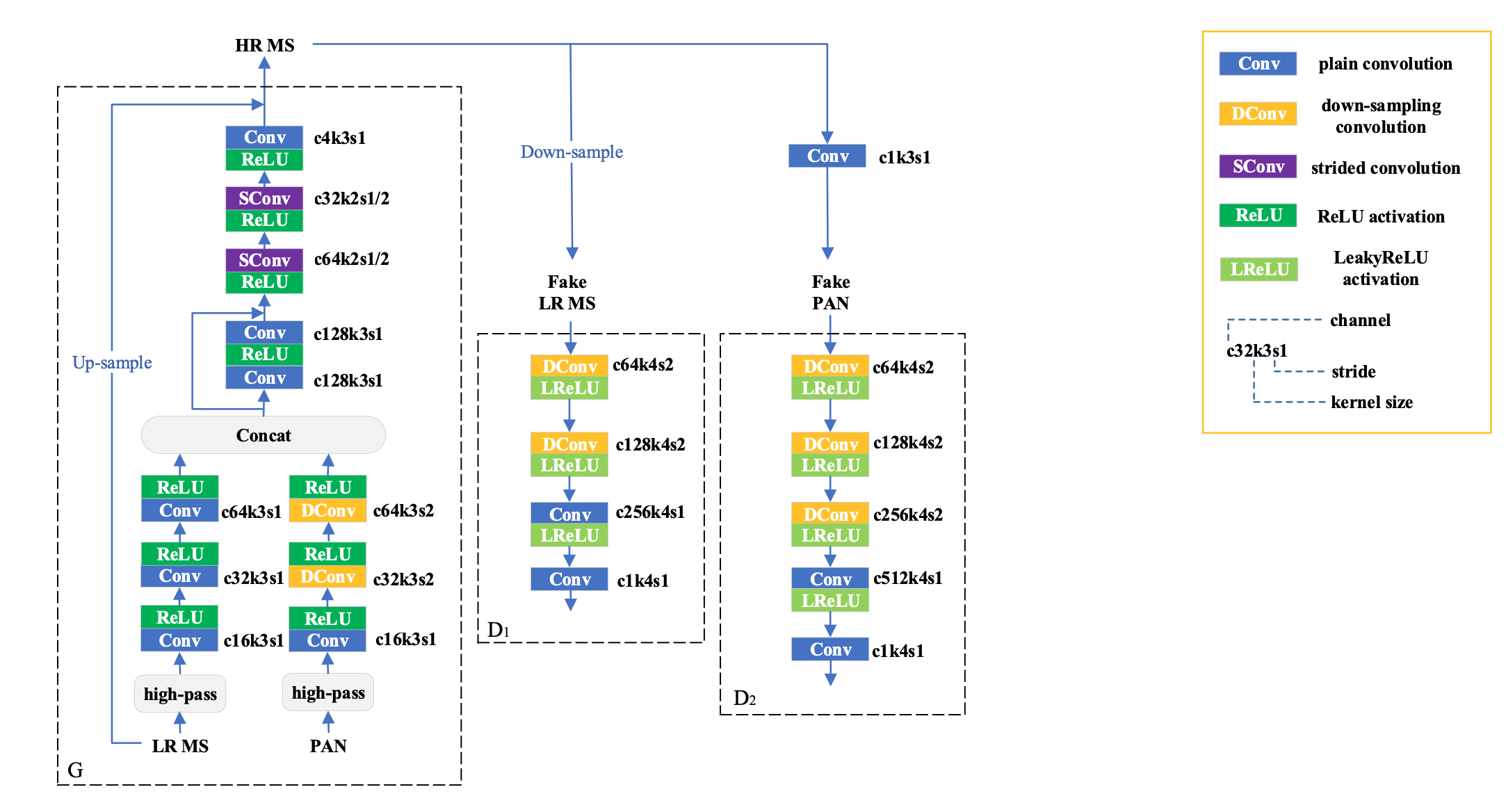}
  \caption{The architecture of our proposed model, PGMAN. The generator takes the original LR MS and PAN images as inputs and generates the HR MS image. The pan-sharpened result will be degraded spatially and spectrally to form a pair of fake inputs for multi-adversarial learning. According to the feedback from these two discriminators, the generator will minimize the distance between the real and fake distributions and further improve the spatial and spectral quality of the fusion results. The parameters of each layer in the neural network are given on the right side.}
  \label{fig::Overview}
\end{figure*}

\section{Method} \label{section::Method}

\subsection{Network Architecture}

\subsubsection{Generator Architecture} 

We design the generator based on the architecture of TFNet~\cite{TFNet} and make the following modifications to it to further improve the quality of the pan-sharpened images. Firstly, inspired by PanNet \cite{PanNet}, the generator is trained on the high-pass domain and the output of it is added to the up-sampled LR MS image for better spectral preservation. Generally, the high-pass domain of images usually contains more spatial details. Moreover, learning the residual between the LR MS image and the final HR MS image can stabilize the training process. Secondly, considering that the input pair PAN and MS images are of different sizes, we build two independent feature extraction (FE) sub-networks. The PAN FE sub-network has two stride-2 convolutions for down-sampling, while the MS FE sub-network has two stride-1 convolutions to maintain the feature map resolution without down-sampling. We concatenate feature maps produced by these two sub-networks and append a residual block \cite{ResNet} to achieve fusion. Finally, two successive fractionally-strided convolutions that both are with a stride of $\frac{1}{2}$ are applied to up-sample the feature maps to meet the size of the desired HR MS. The outputs are high-frequency parts of the pan-sharpened MS images. We add them to the up-sampled LR MS images to obtain the final results. Fig.~\ref{fig::Overview} shows the detailed architecture of our generator.

\subsubsection{Discriminators Architecture}

We use two discriminators for verifying the consistency in pan-sharpening process. Firstly, we down-sample the fused images to the same spatial resolution as the LR MS images and then apply the discriminator-1 $D_1$ to enforce them to have the same spectral information. Secondly, the discriminator-2 $D_2$ is applied to match the spatial structures of the fused images to that of the PAN images. Different from \cite{Pan-GAN}, we don't use the global average pooling or maximum pooling to obtain the spectrally degraded version of the fused HR MS. Instead, we train an auxiliary network consisting of only one $3 \times 3$ convolution layer to estimate the transformation from MS images to PAN images. The auxiliary spectral degradation network is pre-trained individually with the LR MS and the down-sampled PAN images and will be fixed when training the generator and discriminators. We use discriminators similar to the one used in \cite{pix2pix}. Since the input LR MS and PAN are with different image sizes and channels, the two discriminators have distinct architectures as shown in Fig.~\ref{fig::Overview}. $D_2$ has one more convolution layer to down-sample the feature maps because the PAN image has a larger image size. As suggested by WGAN-GP\cite{WGAN-GP}, we remove the last activation function and batch normalization layers in our discriminators. 

As can be seen, our generator and discriminators are fully convolutional, which makes our model easy to train and can accept PAN and MS images with arbitrary sizes in the testing phase.

\begin{table}[t]
\centering
  \caption{List of notations used in this paper.}
  \label{tab::notation}
  \begin{tabular}{cl}
    \toprule
    Notation & Description \\
    \midrule
    $P$ & the pan-sharpened image from the generator \\
    \midrule
    $X$ & the input multi-spectral image \\
    \midrule
    $Y$ & the input panchromatic image \\
    \midrule
    $K$ & the number of bands of a MS image \\
    \midrule
    $P_i$ & the $i$-th band of the image P \\
    \midrule
    $\tilde{P} $ & a spatially degraded version of the image P \\
    \midrule
    $\hat{P} $ & a spectrally degraded version of the image P \\
    \midrule
    $ N $ & the number of a batch \\
    \midrule
    $ n $ & the id of a sample \\
    \midrule
    $ \alpha, \beta, \lambda $  & hyperparameters \\
    \midrule
    $ GP $ & the gradient penalty for discriminators \\
  	\bottomrule
\end{tabular}
\end{table}

\begin{table*}[t]
\centering
  \caption{Details of the datasets used in our experiments}
  \label{tab::dataset}
  \begin{tabular}{ccccrcc}
    \toprule
    \textbf{Sensor} & \textbf{MS bands} & \textbf{bit depth} & \textbf{$\sharp$Images} & \textbf{Resolution (PAN/MS)} & \textbf{Training} & \textbf{Testing} \\
    \midrule
    \multirow{8}{*}{GaoFen-2} & \multirow{8}{*}{4} & \multirow{8}{*}{10} & \multirow{8}{*}{7}
    & \multirow{4}{*}{Reduced(3.2/12.8 m)} & \textbf{$\sharp$Patches:} 24000 & \textbf{$\sharp$Patches:} 286 \\
    & & & & & \textbf{input PAN size:} 256 $\times$ 256 $\times$ 1 & \textbf{input PAN size:} 400 $\times$ 400 $\times$ 1 \\
    & & & & & \textbf{input LRMS size:} 64 $\times$ 64 $\times$ 4 & \textbf{input LRMS size:} 100 $\times$ 100 $\times$ 4 \\
    & & & & & \textbf{output size:} 256 $\times$ 256 $\times$ 4 & \textbf{output size:} 400 $\times$ 400 $\times$ 4 \\
    \cline{5-7}
    & & & & \multirow{4}{*}{Full (0.8/3.2 m)}  & \textbf{$\sharp$Patches:} 24000 & \textbf{$\sharp$Patches:} 286 \\
    & & & & & \textbf{input PAN size:} 256 $\times$ 256 $\times$ 1 & \textbf{input PAN size:} 400 $\times$ 400 $\times$ 1  \\
    & & & & & \textbf{input LRMS size:} 64 $\times$ 64 $\times$ 4 & \textbf{input LRMS size:} 100 $\times$ 100 $\times$ 4 \\
    & & & & & \textbf{output size:} 256 $\times$ 256 $\times$ 4 &  \textbf{output size:} 400 $\times$ 400 $\times$ 4 \\
    \midrule
    \multirow{8}{*}{QuickBird} & \multirow{8}{*}{4} & \multirow{8}{*}{11} & \multirow{8}{*}{5}
    & \multirow{4}{*}{Reduced (2.4/9.6 m)} & \textbf{$\sharp$Patches:} 16000 & \textbf{$\sharp$Patches:} 40 \\
    & & & & & \textbf{input PAN size:} 256 $\times$ 256 $\times$ 1 & \textbf{input PAN size:} 400 $\times$ 400 $\times$ 1 \\
    & & & & & \textbf{input LRMS size:} 64 $\times$ 64 $\times$ 4 & \textbf{input LRMS size:} 100 $\times$ 100 $\times$ 4 \\
    & & & & & \textbf{output size:} 256 $\times$ 256 $\times$ 4 & \textbf{output size:} 400 $\times$ 400 $\times$ 4 \\
    \cline{5-7}
    & & & & \multirow{4}{*}{Full (0.6/2.4 m)}  & \textbf{$\sharp$Patches:} 16000 & \textbf{$\sharp$Patches:} 828 \\
    & & & & & \textbf{input PAN size:} 256 $\times$ 256 $\times$ 1 & \textbf{input PAN size:} 400 $\times$ 400 $\times$ 1  \\
    & & & & & \textbf{input LRMS size:} 64 $\times$ 64 $\times$ 4 & \textbf{input LRMS size:} 100 $\times$ 100 $\times$ 4 \\
    & & & & & \textbf{output size:} 256 $\times$ 256 $\times$ 4 &  \textbf{output size:} 400 $\times$ 400 $\times$ 4 \\
    \midrule
    \multirow{8}{*}{WorldView-3} & \multirow{8}{*}{8} & \multirow{8}{*}{11} & \multirow{8}{*}{3}
    & \multirow{4}{*}{Reduced (1.6/6.4 m)} & \textbf{$\sharp$Patches:} 10000 & \textbf{$\sharp$Patches:} 99 \\
    & & & & & \textbf{input PAN size:} 256 $\times$ 256 $\times$ 1 & \textbf{input PAN size:} 400 $\times$ 400 $\times$ 1 \\
    & & & & & \textbf{input LRMS size:} 64 $\times$ 64 $\times$ 4 & \textbf{input LRMS size:} 100 $\times$ 100 $\times$ 4 \\
    & & & & & \textbf{output size:} 256 $\times$ 256 $\times$ 4 & \textbf{output size:} 400 $\times$ 400 $\times$ 4 \\
    \cline{5-7}
    & & & & \multirow{4}{*}{Full (0.4/1.6 m)}  & \textbf{$\sharp$Patches:} 10000 & \textbf{$\sharp$Patches:} 1960 \\
    & & & & & \textbf{input PAN size:} 256 $\times$ 256 $\times$ 1 & \textbf{input PAN size:} 400 $\times$ 400 $\times$ 1  \\
    & & & & & \textbf{input LRMS size:} 64 $\times$ 64 $\times$ 4 & \textbf{input LRMS size:} 100 $\times$ 100 $\times$ 4 \\
    & & & & & \textbf{output size:} 256 $\times$ 256 $\times$ 4 &  \textbf{output size:} 400 $\times$ 400 $\times$ 4 \\
  	\bottomrule
\end{tabular}
\end{table*}

\subsection{Loss Function}

For simplification and convenience, Table~\ref{tab::notation} lists some key notations used in the following of this paper.

\subsubsection{Q-Loss}
Supervised learning-based methods usually employ $L_1$ or $L_2$ loss to train the networks. However, under the unsupervised setting, there are no ideal images to be compared with. In this work, we attempt to devise an alternative solution that can quantify the quality of the fusion result with reference to the inputs rather than the ground truth. The intuition behind this is that there must be some consistencies across modalities, which means there is a measure that we can obtain in LR MS and PAN images and it still holds when applying it to the HR pan-sharpened MS images. Image quality index (QI) \cite{QIndex} provides a statistical similarity measurement between two monochrome images. To measure the spectral consistency, we can calculate the QI values between any couple of spectral bands in the LR MS image and compare them with those in the pan-sharpened image. Analogously, the QI values between each spectral band in the MS image and the PAN image should be consistent with the QI values between each spectral band in the pan-sharpened image and the PAN image, which defines the spatial consistency. The rationale behind it is, when the spectral information is translated from the coarse-scale to the fine-scale in spatial resolution, the QI values should be unchanged after fusion.

Recall that, in the pan-sharpening paradigm, the inter-relationships measured by QI between any couple of spectral bands of MS images should be unchanged after fusion, otherwise the pan-sharpened MS images may have spectral distortions. Furthermore, the inter-relationships between each band of the MS and the same sized PAN image should be preserved across scales.  Therefore, the spatial and spectral consistencies can be directly calculated from the pan-sharpened image, the LR MS image, and the PAN image without the ground truth. The underlying assumption of inter-relationships consistency of cross-similarity is demonstrated by the fact that the true high-resolution MS data, whenever available, exhibit spectral and spatial distortions that are both zero, within the approximations of the model, and definitely lower than those attained by any fusion method. To describe this, quantitatively, we introduce a non-GT loss on top of QNR \cite{QNR}, which defined as follows: 

\begin{equation}
	\mathcal{L}_{Q} = 1 - QNR
\end{equation}

QNR is the abbreviation of \emph{Quality with No Reference}, which is a combination of $D_\lambda$ and $D_s$. $D_\lambda$ is a spectral distortion index while $D_s$ is a spatial quality metric complementary to $D_\lambda$.

\begin{equation}
	QNR = (1 - D_\lambda) (1 - D_s)
\end{equation}

The optimal value of $QNR$ is 1. 

\begin{equation}
	D_\lambda = \sqrt{
	\frac{2}{K(K-1)} \sum_{i=1}^K \sum_{\underbrace{j=1}_{i\neq j}}^K
	|Q(P_i, P_j) - Q(X_i, X_j) |
	}
\end{equation}
where $P$ is the pan-sharpened result, $X$ stands for the LR MS input and $K$ is the number of bands. $P_i$ and $X_i$ represent the $i$th band of them, respectively. $Q$ stands for Image quality index (QI) \cite{QIndex}. It is defined as follows:

\begin{equation}
	Q(x, y) = \frac{4 \sigma_{xy} \cdot \bar{x} \cdot \bar{y} }{
	( \sigma_x^2 + \sigma_y^2 ) ( \bar{x}^2 + \bar{y}^2 )
	}
\end{equation}
where $\sigma_{xy}$ means the covariance between $x$ and $y$, and $\sigma^2_{x}$ and $\sigma^2_{y}$ are the variances of $x$ and $y$, respectively. $\bar{x}$ and $\bar{y}$ are the means of $x$ and $y$, respectively.

\begin{equation}
	D_s = \sqrt{
	\frac{1}{K} \sum_{i=1}^K | Q(P_i, Y) - Q(X_i, \tilde{Y}) |
	}
\end{equation}
where $Y$ is the panchromatic input and $\tilde{Y}$ is the degraded low- resolution version of it.

This loss function enables us to measure the qualities of the fused images from input PAN and MS images without ground truth HR MS images.

\subsubsection{Adversarial Loss}

We design two discriminators, $D_1$ for spectral preservation and $D_2$ for spatial preservation. The generator learns to preserve more spectral information to cheat $D_1$ which is able to distinguish the real and fake LR MS images and preserve more spatial details to cheat $D_2$ which is able to distinguish the real PAN image and the spectral degraded fusion result. The loss function of the generator $G$ takes the form of:
\begin{equation}\label{eq:adversarial_loss}
\begin{aligned}
	\mathcal{L}_G = \frac{1}{N} \sum_{n=1}^N - \alpha D_1(\tilde{P}^{(n)}) - \beta D_2(\hat{P}^{(n)} ) \\
	 + \mathcal{L}_{Q} (P^{(n)}, X^{(n)}, Y^{(n)})
\end{aligned}
\end{equation}
where $N$ is the number of samples. $P$, $X$, $Y$ are pan-sharpened MS, LR MS, and PAN images, respectively. $\tilde{P}$ and $\hat{P}$ stand for the spatially and spectrally degraded $P$. $\alpha$ and $\beta$ are hyper-parameters.

To stabilize the training, we employ WGAN-GP \cite{WGAN-GP} as a basic framework, $i.e.$, using Wasserstein distance \cite{WGAN} and applying gradient penalty \cite{WGAN-GP} to discriminators. The loss functions of $D_1$ and $D_2$ are formulated as follows:

\begin{equation}\label{eq:d1_loss}
\begin{aligned}
	\mathcal{L}_{D_1} = \frac{1}{N} \sum_{n=1}^N - D_1(X^{(n)}) + D_1(\tilde{P}^{(n)}) \\ 
	+ \lambda GP(D_1, X^{(n)}, \tilde{P}^{(n)})
\end{aligned}
\end{equation}

\begin{equation}\label{eq:d2_loss}
\begin{aligned}
	\mathcal{L}_{D_2} = \frac{1}{N} \sum_{n=1}^N - D_2(Y^{(n)}) + D_2(\hat{P}^{(n)}) \\ 
	+ \lambda GP(D_2, Y^{(n)}, \hat{P}^{(n)})
\end{aligned}
\end{equation}

where $GP$ is the gradient penalty for discriminators and $\lambda$ is a hyper-parameter. 

\subsection{Training Details}

Our method is implemented in PyTorch \cite{PyTorch} and trained on a single NVIDIA Titan 1080Ti GPU. The batch size and learning rate are set as 8 and 1e-4, respectively. The hyper-parameters in Eqs.~\ref{eq:adversarial_loss}, \ref{eq:d1_loss} and \ref{eq:d2_loss} are set as $\alpha = 2e-4, \beta = 1e-4$, and $\lambda = 100$. Adam optimizer \cite{Adam} is used to train the model for 20 epochs with fixed hyper-parameters $\beta_1 =0$ and $\beta_2=0.9$. It should be noted that, we process the images at the raw bit depth in both training and testing phases without normalizing them into displayable 8-bit images.  

\section{Experiments and Results} \label{section::Experiments and Results}

\subsection{Datasets}

We conduct extensive experiments on three datasets with images collected from GaoFen-2, QuickBird and WorldView-3 satellites. The detailed information of the datasets can be seen in Table~\ref{tab::dataset}. For comparison between the supervised and unsupervised methods, we build the datasets under both reduced-scale (based on Wald's protocol~\cite{wald}) and full-scale settings. Wald's protocol has been widely used for assessment of pan-sharpening methods, in which the original MS and PAN images are spatially degraded before feeding into models, the reducing factor being the ratio between their spatial resolutions, and the original MS images are considered as reference images for comparison. Same as the previous works\cite{PUCNN, Pan-GAN}, we implement it by blurring the full-resolution datasets using a Gaussian filter and then downsampling them with a scaling factor of 4. Under Wald's protocol, the supervised models can be trained on reduced-resolution images using the original MS images as labels. However, under full-resolution setting, there are no reference images so that only unsupervised models can be trained with the full-resolution images as inputs. Although the training environment is limited related to the type of models, testing is without constraints. We can test all models on both reduced-scale and full-scale images whether they need supervised labels or not. 

Moreover, because of the very large size of remote sensing images, for example, PAN and MS images of GaoFen-2 are with the size of about $30,000\times30,000$ and $7,500\times7,500$ pixels, respectively, which is too large to feed into a neural network, we crop these images into small patches to form training and testing sets. As for the testing set, we crop the remote sensing images orderly in small overlapping regions between neighborhood patches to remove border effects as is the common practice. It is noted that there is no overlapping between the training and testing sets. 

\begin{table*}[t]
\centering
  \caption{Quantity results on GaoFen-2 dataset. The best result in each group is in \textbf{bold} font. The last row indicates the ideal value of each metric.}
  \label{tab::GF-2 result}
  \begin{tabular}{r|r|c|c|c|c|c|c}
    \toprule
    \multirow{2}{*}{Type} & \multirow{2}{*}{Model}	& \multicolumn{3}{c|}{Non-reference Metrics} & \multicolumn{3}{c}{Reference Metrics} \\
     & & \textbf{D$_\lambda$} 	& \textbf{D$_s$}		& \textbf{QNR} 		& \textbf{SAM} 		& \textbf{ERGAS} 	& \textbf{SSIM} \\
    \midrule
    \multirow{7}{*}{Traditional} & BDSD \cite{BDSD} & 0.0283$\pm$0.0184 & 0.0451$\pm$0.0214 & 0.9282$\pm$0.0370 & 3.0869$\pm$0.3199 & 6.2093$\pm$0.6545 & 0.6969$\pm$0.0266 \\
    & GS \cite{GS} & 0.0297$\pm$0.0246 & 0.0475$\pm$0.0316 & 0.9249$\pm$0.0513 & 1.9673$\pm$0.2051 & 3.7190$\pm$0.3750 & \textbf{0.8383$\pm$0.0181} \\
    & IHS \cite{IHS} & 0.0281$\pm$0.0224 & 0.0451$\pm$0.0316 & 0.9287$\pm$0.0497 & 2.1299$\pm$0.2151 & 3.8587$\pm$0.4022 & 0.8377$\pm$0.0183 \\
   	& Brovey \cite{Brovey} & 0.0246$\pm$0.0168 & 0.0505$\pm$0.0302 & 0.9265$\pm$0.0418 & \textbf{1.7377$\pm$0.2054} & \textbf{3.4325$\pm$0.3198} & 0.8293$\pm$0.0172 \\
    & HPF \cite{HPFandHPFC} & 0.0204$\pm$0.0116 & \textbf{0.0182$\pm$0.0079} & 0.9618$\pm$0.0175 & 2.0829$\pm$0.2312 & 4.1046$\pm$0.4477 & 0.8097$\pm$0.0199 \\
    & LMM \cite{LMVMandLMM} & 0.0221$\pm$0.0138 & 0.0250$\pm$0.0139 & 0.9537$\pm$0.0256 & 1.8137$\pm$0.2233 & 6.3369$\pm$0.6827 & 0.6397$\pm$0.0274 \\
    & SFIM \cite{SFIM} & \textbf{0.0178$\pm$0.0113} & 0.0193$\pm$0.0074 & \textbf{0.9634$\pm$0.0166} & 1.9509$\pm$0.1881 & 4.2613$\pm$0.4855 & 0.8085$\pm$0.0212 \\ 
    \midrule
    \multirow{5}{*}{Supervised} & PNN \cite{PNN} & 0.0147$\pm$0.0106 & \textbf{0.0313$\pm$0.0144} & \textbf{0.9544$\pm$0.0149} & 1.2717$\pm$0.1157 & 1.4982$\pm$0.1995 & 0.9626$\pm$0.0064\\
    & DRPNN \cite{DRPNN} & 0.0168$\pm$0.0087 & 0.0695$\pm$0.0294 & 0.9150$\pm$0.0308 & 1.2162$\pm$0.1203 & 1.2838$\pm$0.1509 & 0.9708$\pm$0.0041 \\
    & MSDCNN \cite{MSDCNN} & 0.0131$\pm$0.0076 & 0.0409$\pm$0.0229 & 0.9465$\pm$0.0232 & 1.4422$\pm$0.1543 & 1.4904$\pm$0.2065 & 0.9629$\pm$0.0064 \\
    & PanNet \cite{PanNet} & 0.0113$\pm$0.0072 & 0.1234$\pm$0.0429 & 0.8669$\pm$0.0458 & 1.0881$\pm$0.1140 & 1.2923$\pm$0.1655 & 0.9712$\pm$0.0046 \\
    & PSGAN \cite{PSGAN} & \textbf{0.0082$\pm$0.0042} & 0.1012$\pm$0.0360 & 0.8914$\pm$0.0373 & \textbf{1.0572$\pm$0.1040} & \textbf{1.2092$\pm$0.1427} & \textbf{0.9745$\pm$0.0038} \\
	\midrule
	\multirow{2}{*}{Unsupervised} & Pan-GAN \cite{Pan-GAN} & 0.0347$\pm$0.0188 & 0.0274$\pm$0.0179 & 0.9392$\pm$0.0342 & 2.6423$\pm$0.2697 & 5.2453$\pm$0.5608 & 0.7318$\pm$0.0242 \\
	& PGMAN &  \textbf{0.0077$\pm$0.0043} & \textbf{0.0134$\pm$0.0107} & \textbf{0.9790$\pm$0.0129} & \textbf{2.0100$\pm$0.1947} & \textbf{2.6394$\pm$0.2978} & \textbf{0.9166$\pm$0.0111} \\
	\midrule
	\multicolumn{2}{c|}{Ideal Value} & 0	 & 0 	& 1	& 0	& 0 	& 1 \\
	\bottomrule
	\end{tabular}
\end{table*}

\begin{table*}[t]
\centering
  \caption{Quantity results on QuickBird dataset. The best result in each group is in \textbf{bold} font. The last row indicates the ideal value of each metric.}
  \label{tab::QB result}
  \begin{tabular}{r|r|c|c|c|c|c|c}
    \toprule
    \multirow{2}{*}{Type} & \multirow{2}{*}{Model}	& \multicolumn{3}{c|}{Non-reference Metrics} & \multicolumn{3}{c}{Reference Metrics} \\
     & & \textbf{D$_\lambda$} 	& \textbf{D$_s$}		& \textbf{QNR} 		& \textbf{SAM} 		& \textbf{ERGAS} 	& \textbf{SSIM} \\
    \midrule
    \multirow{7}{*}{Traditional} & BDSD \cite{BDSD} & 0.0110$\pm$0.0000 & \textbf{0.0202$\pm$0.0003} & \textbf{0.9690$\pm$0.0004} & 2.0508$\pm$0.1499 & 3.7505$\pm$0.5481 & 0.7940$\pm$0.0028 \\
    & GS \cite{GS} & 0.0202$\pm$0.0003 & 0.0433$\pm$0.0011 & 0.9377$\pm$0.0018 & \textbf{1.5325$\pm$0.0631} & 2.7242$\pm$0.3269 & \textbf{0.8684$\pm$0.0019} \\
    & IHS \cite{IHS} & 0.0222$\pm$0.0002 & 0.0476$\pm$0.0012 & 0.9315$\pm$0.0017 & 1.8226$\pm$0.0624 & 2.8254$\pm$0.3321 & 0.8605$\pm$0.0019 \\
    & Brovey \cite{Brovey} & 0.0226$\pm$0.0002 & 0.0435$\pm$0.0010 & 0.9353$\pm$0.0018 & 1.5730$\pm$0.0291 & \textbf{2.7190$\pm$0.2970} & 0.8600$\pm$0.0019 \\
    & HPF \cite{HPFandHPFC} & 0.0094$\pm$0.0000 & 0.0429$\pm$0.0007 & 0.9481$\pm$0.0008 & 1.7227$\pm$0.0668 & 3.1126$\pm$0.2483 & 0.8489$\pm$0.0011 \\
    & LMM \cite{LMVMandLMM} & 0.0105$\pm$0.0000 & 0.0389$\pm$0.0006 & 0.9510$\pm$0.0006 & 1.6070$\pm$0.0320 & 2.9798$\pm$0.2548 & 0.8367$\pm$0.0015 \\
    & SFIM \cite{SFIM} & \textbf{0.0088$\pm$0.0000} & 0.0441$\pm$0.0008 & 0.9476$\pm$0.0008 & 1.5982$\pm$0.0460 & 3.1296$\pm$0.2503 & 0.8496$\pm$0.0011 \\ 
    \midrule
    \multirow{5}{*}{Supervised} & PNN \cite{PNN} & 0.0467$\pm$0.0004 & 0.0927$\pm$0.0004 & 0.8652$\pm$0.0011 & 2.5468$\pm$0.0887 & 3.3448$\pm$0.1668 & 0.9020$\pm$0.0005 \\
    & DRPNN \cite{DRPNN} & 0.0182$\pm$0.0001 & 0.0500$\pm$0.0003 & 0.9329$\pm$0.0006 & 2.1292$\pm$0.0765 & 2.2699$\pm$0.1625 & 0.9235$\pm$0.0005 \\ 
    & MSDCNN \cite{MSDCNN} & 0.0654$\pm$0.0004 & 0.1116$\pm$0.0009 & 0.8307$\pm$0.0018 & 3.1420$\pm$0.0617 & 3.2128$\pm$0.1386 &  0.8897$\pm$0.0008 \\
    & PanNet \cite{PanNet} & \textbf{0.0048$\pm$0.0000} & \textbf{0.0409$\pm$0.0002} & \textbf{0.9545$\pm$0.0003} & \textbf{1.4493$\pm$0.0683} & \textbf{1.7392$\pm$0.1650} & \textbf{0.9465$\pm$0.0003} \\
    & PSGAN \cite{PSGAN} & 0.0297$\pm$0.0001 & 0.0668$\pm$0.0005 & 0.9055$\pm$0.0007 & 2.1482$\pm$0.0530 & 2.1682$\pm$0.0808 & 0.9266$\pm$0.0002 \\
	\midrule
	\multirow{2}{*}{Unsupervised} & Pan-GAN \cite{Pan-GAN} & 0.0082$\pm$0.0046 & 0.0286$\pm$0.0177 & 0.9633$\pm$0.0173 & 2.6317$\pm$0.3126 & \textbf{2.2248$\pm$0.2379} & \textbf{0.9192$\pm$0.0125} \\
	& PGMAN & \textbf{0.0037$\pm$0.0024} & \textbf{0.0202$\pm$0.0087} & \textbf{0.9762$\pm$0.0093} & \textbf{1.7902$\pm$0.1838} & 2.2581$\pm$0.3308 & 0.9076$\pm$0.0182 \\
	\midrule
	\multicolumn{2}{c|}{Ideal Value} & 0	 & 0 	& 1	& 0	& 0 	& 1 \\
	\bottomrule
	\end{tabular}
\end{table*}

\begin{table*}[t]
\centering
  \caption{Quantity results on WoldView-3 dataset. The best result in each group is in \textbf{bold} font. The last row indicates the ideal value of each metric.}
  \label{tab::WV-3 result}
  \begin{tabular}{r|r|c|c|c|c|c|c}
    \toprule
    \multirow{2}{*}{Type} & \multirow{2}{*}{Model}	& \multicolumn{3}{c|}{Non-reference Metrics} & \multicolumn{3}{c}{Reference Metrics} \\
     & & \textbf{D$_\lambda$} 	& \textbf{D$_s$}		& \textbf{QNR} 		& \textbf{SAM} 		& \textbf{ERGAS} 	& \textbf{SSIM} \\
    \midrule
    \multirow{7}{*}{Traditional} & BDSD \cite{BDSD} & \textbf{0.0260$\pm$0.0321} & \textbf{0.0733$\pm$0.0641} & \textbf{0.9044$\pm$0.0835} & 5.4941$\pm$0.9020 & \textbf{5.5417$\pm$0.6988} & \textbf{0.8493$\pm$0.0274} \\
    & GS \cite{GS} & 0.0461$\pm$0.0525 & 0.1087$\pm$0.0796 & 0.8542$\pm$0.1150 & 5.9630$\pm$0.5248 & 6.2308$\pm$0.6185 & 0.8309$\pm$0.0246 \\
    & IHS \cite{IHS} & 0.0574$\pm$0.0642 & 0.1121$\pm$0.0849 & 0.8421$\pm$0.1273 & 6.2706$\pm$0.7805 & 6.3862$\pm$0.6681 & 0.8199$\pm$0.0269 \\
    & Brovey \cite{Brovey} & 0.0429$\pm$0.0418 & 0.1042$\pm$0.0708 & 0.8601$\pm$0.0979 & 5.6566$\pm$0.9382 & 6.2390$\pm$0.6842 & 0.8338$\pm$0.0254 \\
    & HPF \cite{HPFandHPFC} & 0.0386$\pm$0.0396 & 0.0804$\pm$0.0574 & 0.8862$\pm$0.0840 & 5.5389$\pm$0.8940 & 6.7569$\pm$1.0258 & 0.8102$\pm$0.0356 \\
    & LMM \cite{LMVMandLMM} & 0.0433$\pm$0.0446 & 0.0759$\pm$0.0575 & 0.8865$\pm$0.0879 & 5.7981$\pm$0.9798 & 6.9729$\pm$1.1321 & 0.8084$\pm$0.0364 \\
    & SFIM \cite{SFIM} & 0.0384$\pm$0.0437 & 0.0754$\pm$0.0512 & 0.8909$\pm$0.0803 & \textbf{5.3298$\pm$0.8431} & 7.3518$\pm$1.9554 & 0.8154$\pm$0.0361 \\ 
    \midrule
    \multirow{5}{*}{Supervised} & PNN \cite{PNN} & 0.0412$\pm$0.0402 & 0.0445$\pm$0.0340 & 0.9171$\pm$0.0633 & 5.2637$\pm$0.5915 & 4.1083$\pm$0.5081 & 0.9250$\pm$0.0191 \\
    & DRPNN \cite{DRPNN} & 0.0478$\pm$0.0352 & 0.0533$\pm$0.0250 & 0.9020$\pm$0.0492 & 6.1427$\pm$0.5190 & 4.5375$\pm$0.3573 & 0.9301$\pm$0.0145 \\ 
    & MSDCNN \cite{MSDCNN} & 0.0415$\pm$0.0424 & 0.0494$\pm$0.0321 & 0.9122$\pm$0.0635 & 5.2105$\pm$0.5746 & 4.2856$\pm$0.5286 & 0.9225$\pm$0.0187 \\
    & PanNet \cite{PanNet} & \textbf{0.0374$\pm$0.0410} & \textbf{0.0355$\pm$0.0327} & \textbf{0.9294$\pm$0.0639} & 4.8066$\pm$0.5473 & 4.3263$\pm$0.5742 & 0.9180$\pm$0.0202 \\
    & PSGAN \cite{PSGAN} & 0.0392$\pm$0.0456 & 0.0534$\pm$0.0335 & 0.9108$\pm$0.0674 & \textbf{4.5523$\pm$0.5049} & \textbf{4.1070$\pm$0.4402} & \textbf{0.9311$\pm$0.0192} \\
	\midrule
	\multirow{2}{*}{Unsupervised} & Pan-GAN \cite{Pan-GAN} & 0.0396$\pm$0.0449 & 0.0638$\pm$0.0529 & 0.9013$\pm$0.0839 & \textbf{5.9902$\pm$1.0484} & 7.1762$\pm$1.2440 & \textbf{0.7857$\pm$0.0434} \\
	& PGMAN & \textbf{0.0129$\pm$0.0128} & \textbf{0.0409$\pm$0.0242} & \textbf{0.9469$\pm$0.0327} & 6.3959$\pm$0.9904 & \textbf{6.8348$\pm$1.2325} & 0.7617$\pm$0.0563 \\
	\midrule
	\multicolumn{2}{c|}{Ideal Value} & 0	 & 0 	& 1	& 0	& 0 	& 1 \\
	\bottomrule
	\end{tabular}
\end{table*}

\subsection{Metrics}

In order to examine the performance of models on both reduced-scale and full-scale images, we use reference and non-reference metrics, respectively.

\subsubsection{Non-reference Metrics}
On full-resolution, there is no ground truth and we use the non-reference metrics, which can be calculated without the target. $D_\lambda$, $D_s$ and QNR are widely used non-reference metrics in pan-sharpening task, which have been introduced in Section \ref{section::Method} to support our loss function.

\subsubsection{Reference Metrics}
Under Wald's protocol \cite{wald}, we can also validate models on reduced-resolution where the PAN and MS images are down-sampled so that the origin MS image is as ground truth. Therefore on low-resolution, we use the reference metrics:

\begin{itemize}
\item \textbf{SAM} \cite{SAM} measures spectral distortions of pan-sharpened images comparing with the reference images. 
\begin{equation}
    SAM(x_1, x_2) = \arccos{(\frac{x_1 \cdot x_2}{||x_1|| \cdot ||x_2||})}
\end{equation}
where $x_1$, $x_2$ are two spectral vectors from the pan-sharpened result and the reference result.

\item \textbf{ERGAS} \cite{ERGAS} The \emph{erreur relative globale adimensionnelle de synth\`ese} (ERGAS), also known as the relative global dimensional synthesis error is a commonly used global quality index. It is given by:
	
\begin{equation} 
\mathrm{ERGAS} = 100\frac{h}{l}\sqrt{\frac{1}{N}\sum^N_{i=1}\left(\frac{\mathrm{RMSE}(B_i)}{M(B_i)}\right)^2}
\end{equation}

where $h$ and $l$ are the spatial resolutions of PAN and MS images; RMSE($B_i$) is the root mean square error between the $i$th band of the fused and reference image; $M(B_i)$ is the mean value of the original MS band $B_i$.	
	
\item \textbf{SSIM} \cite{SSIM} is a widely used metric which models the loss and distortion according to the similarities in light, contrast, and structure information.
\begin{equation}
    SSIM(x, y) = \frac{
    (2 \mu_x \mu_y + c_1) (2 \sigma_{xy} + c_2)
    }{
    (\mu_x^2 + \mu_y^2 + c_1) (\sigma_x^2 + \sigma_y^2 + c_2)
    }
\end{equation}
where $\sigma_{xy}$ means the covariance between $x$ and $y$, and $\sigma^2_{x}$ and $\sigma^2_{y}$ are the variances of $x$ and $y$, respectively. $\mu_{x}$ and $\mu_{y}$ are the means of $x$ and $y$, respectively. $c_1$ and $c_2$ are fixed constants.

\end{itemize}

\subsection{Comparison with State-of-the-arts}

Fourteen methods including seven traditional methods, five supervised methods, and two unsupervised methods are employed for comparison: 

\textbf{Traditional methods} include BDSD \cite{BDSD}, GS \cite{GS}, IHS \cite{IHS}, Brovey \cite{Brovey}, HPF \cite{HPFandHPFC}, LMM \cite{LMVMandLMM} and SFIM \cite{SFIM}.

\textbf{Supervised methods} are PNN \cite{PNN}, DRPNN \cite{DRPNN}, MSDCNN \cite{MSDCNN}, PanNet \cite{PanNet} and PSGAN \cite{PSGAN}. These methods are state-of-the-art supervised deep learning based pan-sharpening methods. The first four models are CNN-based models and the last one is based on GAN. 

\textbf{Unsupervised methods} include Pan-GAN \cite{Pan-GAN} and our proposed PGMAN. 

\subsection{Quantitative Results}

For non-reference metrics, D$_\lambda$ is used to examine the spectral distortions in full-scale images, D$_s$ is for the spatial distortions, and QNR is a comprehensive metric. For reference metrics, SAM is used to examine the spectral distortions in reduced-scale images, ERGAS is for the spatial distortions, and SSIM is a comprehensive metric.

Table~\ref{tab::GF-2 result} shows the quantitative assessments on GaoFen-2. The best value in each group is in \textbf{bold} font. For the non-reference metrics, our proposed PGMAN takes the advantage of the unsupervised learning and obtains the best average values on D$_\lambda$, D$_s$ and QNR, which is much better than all the other competitors. As for the reference metrics, unsupervised methods fall behind supervised methods because they do not utilize the label information while the supervised methods can directly minimize the difference between the output pan-sharpened images and the ground truth. Nevertheless, our proposed PGMAN still obtains the best results among the unsupervised methods on the reference metrics. PGMAN also achieves better values than all of the traditional methods in terms of ERGAS and SSIM.

Table~\ref{tab::QB result} shows the quantitative assessments on QuickBird. Similar to the results of GaoFen-2 dataset, here, our proposed model, PGMAN maintains its superiority and achieves the best values on all non-reference metrics, which exceed the performance of other traditional and supervised methods. PGMAN also obtains the best value in terms of SAM in the unsupervised methods, while is a little behind Pan-GAN on ERGAS and SSIM.

And we also conduct an evaluation on the 8-band images from WorldView-3 dataset. Table~\ref{tab::WV-3 result} shows the quantitative assessments on it. The results show that our proposed PGMAN still maintains its superiority and achieves the best values on all non-reference metrics, showing the effectiveness and the great practical value of it. For the reference metrics, both unsupervised methods fall a lot. In 8-band WV3 dataset, the generalization ability from the full-scale dataset to the reduced-scale dataset may be affected due to the added spectral bands. It remains a challenge to the unsupervised methods trained in the full-scale dataset.

\begin{table*}[t]
\centering
  \caption{Ablation study of the loss function on GaoFen-2 dataset. ``$\surd$" means including the referred loss item. The last row indicates the ideal value of each metric.}
  \label{tab::ablation results}
  \begin{tabular}{c|c|c|c|c|c|c|c}
    \toprule
    \multicolumn{2}{c|}{Loss item} & \multicolumn{3}{c|}{Non-reference Metrics} & \multicolumn{3}{c}{Reference Metrics} \\
    ~~$\mathcal{L}_{Q}$~~ & $\mathcal{L}_{adv}$ & \textbf{D$_\lambda$} & \textbf{D$_s$} & \textbf{QNR} & \textbf{SAM} 	& \textbf{ERGAS} 	& \textbf{SSIM} \\
    \midrule     
    & $\surd$ & 0.0072$\pm$0.0036 & 0.1117$\pm$0.0369 & 0.8819$\pm$0.0377 & 2.3673$\pm$0.2382 & 3.4380$\pm$0.4005 & 0.8375$\pm$0.0228 \\
    $\surd$ & & 0.0080$\pm$0.0061 & 0.0182$\pm$0.0119 & 0.9740$\pm$0.0162 & 2.1384$\pm$0.1838 & 2.8171$\pm$0.3070 & 0.9074$\pm$0.0108 \\
    $\surd$ & $\surd$ & 0.0077$\pm$0.0043 & 0.0134$\pm$0.0107 & 0.9790$\pm$0.0129 & 2.0100$\pm$0.1947 & 2.6394$\pm$0.2978 & 0.9166$\pm$0.0111 \\
	\midrule
	\multicolumn{2}{c|}{Ideal Value} & 0	 & 0 	& 1	& 0	& 0 & 1 \\
    \bottomrule
\end{tabular}
\end{table*}

\begin{table*}[t]
\centering
  \caption{Parameter analysis on GaoFen-2 dataset.}
  \label{tab::parameter analysis}
  \begin{tabular}{c|c|c|c|c|c|c|c}
    \toprule
    \multicolumn{2}{c|}{Parameter} & \multicolumn{3}{c|}{Non-reference Metrics} & \multicolumn{3}{c}{Reference Metrics} \\
    ~~$\mathcal{\alpha}$~~ & $\mathcal{\beta}$ & \textbf{D$_\lambda$} & \textbf{D$_s$} & \textbf{QNR} & \textbf{SAM} 	& \textbf{ERGAS} 	& \textbf{SSIM} \\
    \midrule     
    1e-3 & 2e-3 & 0.0080$\pm$0.0045 & 0.0136$\pm$0.0085 & 0.9786$\pm$0.0107 & 2.1021$\pm$0.1727 & 2.7228$\pm$0.2840 & 0.9112$\pm$0.0111 \\
    2e-3 & 1e-3 & 0.0074$\pm$0.0040 & 0.0222$\pm$0.0088 & 0.9706$\pm$0.0094 & 1.9444$\pm$0.1955 & 2.4874$\pm$0.2799 & 0.9130$\pm$0.0112 \\
    1e-4 & 2e-4 & 0.0087$\pm$0.0036 & 0.0144$\pm$0.0102 & 0.9770$\pm$0.0112 & 2.0712$\pm$0.1856 & 2.7020$\pm$0.2962 & 0.9135$\pm$0.0114 \\
    2e-4 & 1e-4 & 0.0077$\pm$0.0043 & 0.0134$\pm$0.0107 & 0.9790$\pm$0.0129 & 2.0100$\pm$0.1947 & 2.6394$\pm$0.2978 & 0.9166$\pm$0.0111 \\
    1e-5 & 2e-5 & 0.0074$\pm$0.0039 & 0.0139$\pm$0.0098 & 0.9788$\pm$0.0115 & 2.2276$\pm$0.1949 & 2.9133$\pm$0.3134 & 0.9080$\pm$0.0115 \\
    0 & 0 & 0.0080$\pm$0.0061 & 0.0182$\pm$0.0119 & 0.9740$\pm$0.0162 & 2.1384$\pm$0.1838 & 2.8171$\pm$0.3070 & 0.9074$\pm$0.0108  \\
	\midrule
	\multicolumn{2}{c|}{Ideal Value} & 0	 & 0 	& 1	& 0	& 0 & 1 \\
    \bottomrule
\end{tabular}
\end{table*}

Generally, traditional models do not rely on any training data and make a medium performance regardless of the kind of testing environment. Supervised methods tend to be better on the reduced-scale images because they make full use of the supervision information and directly minimize the loss between the predicted pan-sharpened and the ground truth images. However, this may make them overfit the reduced-scale images and generalize poorly on full-resolution test images, and sometimes these methods even perform worse than traditional models. The best non-reference metrics achieved by supervised methods are PNN (GaoFen-2 dataset, Table~\ref{tab::GF-2 result}) and PanNet (QuickBird dataset, Table~\ref{tab::QB result}), the reason may be that they consist of fewer parameters (see Table~\ref{tab::efficiency study}) and thus with lower possibility to become overfitting. Here, unsupervised models take the advantages of learning from the full-scale images and obtain the best non-reference metrics. Furthermore, our proposed model, PGMAN, surpasses all other unsupervised methods. Our method significantly improves the results, which indicates that full-resolution images indeed provide rich spatial and spectral information and are helpful for improving the quality of the pan-sharpened images. Although our method falls behind the supervised methods when testing on the reduced-scale images, considering that there is no ground truth used in our method, the results are quite promising. And more importantly, the proposed PGMAN obtains the best QNR when applied to full-resolution images, showing the great practical value of it. 

\begin{figure*}[t]
\centering

\begin{minipage}[b]{.23\linewidth}
  \centering
  \includegraphics[width=\textwidth]{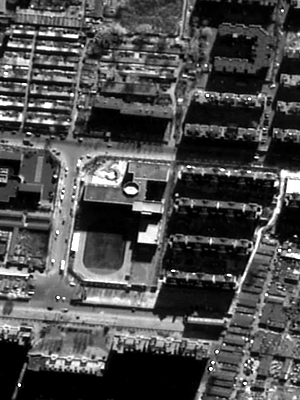}
  \centerline{(a) PAN}\medskip
\end{minipage}
\begin{minipage}[b]{.23\linewidth}
  \centering
  \includegraphics[width=\textwidth]{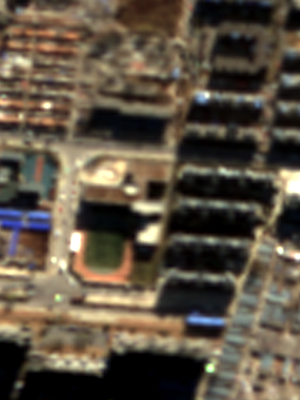}
  \centerline{(b) LR MS}\medskip
\end{minipage}
\begin{minipage}[b]{.23\linewidth}
  \centering
  \includegraphics[width=\textwidth]{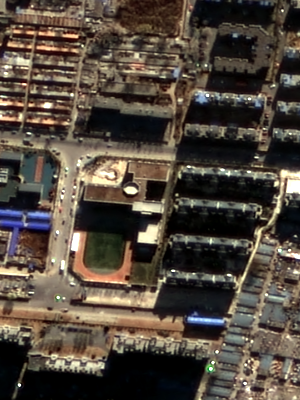}
  \centerline{(c) BDSD \cite{BDSD}}\medskip
\end{minipage}
\begin{minipage}[b]{.23\linewidth}
  \centering
  \includegraphics[width=\textwidth]{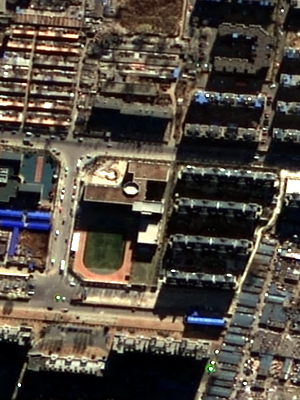}
  \centerline{(d) GS \cite{GS}}\medskip
\end{minipage}

\begin{minipage}[b]{.23\linewidth}
  \centering
  \includegraphics[width=\textwidth]{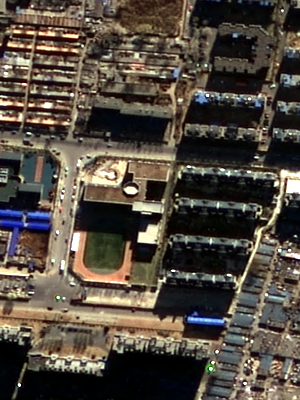}
  \centerline{(e) IHS \cite{IHS}}\medskip
\end{minipage}
\begin{minipage}[b]{.23\linewidth}
  \centering
  \includegraphics[width=\textwidth]{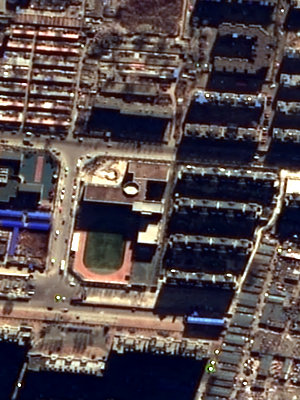}
  \centerline{(f) Brovey \cite{Brovey} }\medskip
\end{minipage}
\begin{minipage}[b]{.23\linewidth}
  \centering
  \includegraphics[width=\textwidth]{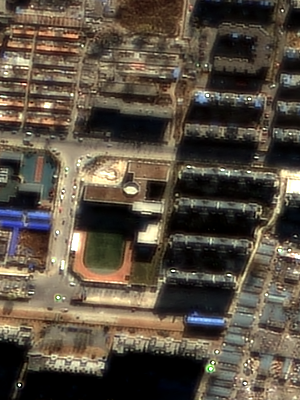}
  \centerline{(g) HPF \cite{HPFandHPFC} }\medskip
\end{minipage}
\begin{minipage}[b]{.23\linewidth}
  \centering
  \includegraphics[width=\textwidth]{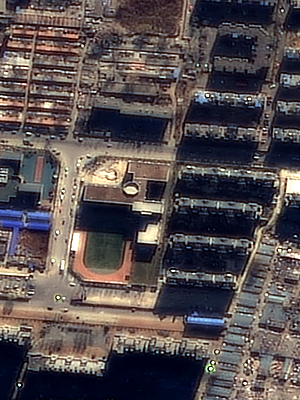}
  \centerline{(h) LMM \cite{LMVMandLMM} }\medskip
\end{minipage}

\begin{minipage}[b]{.23\linewidth}
  \centering
  \includegraphics[width=\textwidth]{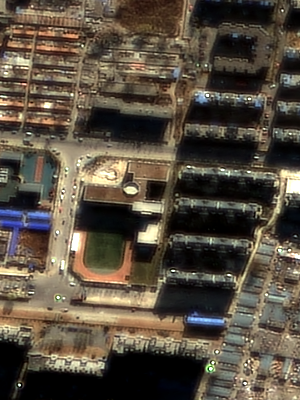}
  \centerline{(i) SFIM \cite{SFIM} }\medskip
\end{minipage}
\begin{minipage}[b]{.23\linewidth}
  \centering
  \includegraphics[width=\textwidth]{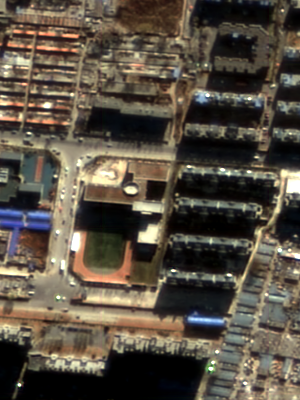}
  \centerline{(j) PNN \cite{PNN} }\medskip
\end{minipage}
\begin{minipage}[b]{.23\linewidth}
  \centering
  \includegraphics[width=\textwidth]{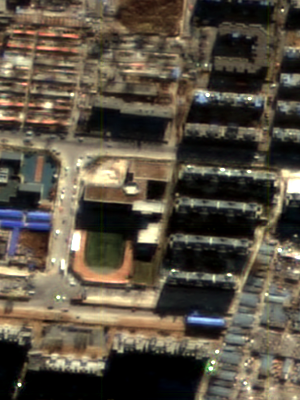}
  \centerline{(k) DRPNN \cite{DRPNN} }\medskip
\end{minipage}
\begin{minipage}[b]{.23\linewidth}
  \centering
  \includegraphics[width=\textwidth]{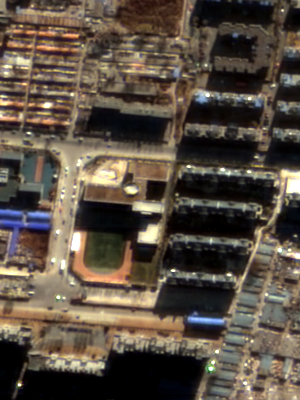}
  \centerline{(l) MSDCNN \cite{MSDCNN} }\medskip
\end{minipage}

\begin{minipage}[b]{.23\linewidth}
  \centering
  \includegraphics[width=\textwidth]{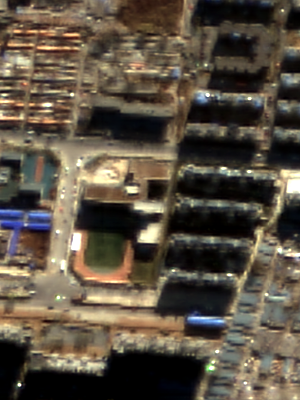}
  \centerline{(m) PanNet \cite{PanNet} }\medskip
\end{minipage}
\begin{minipage}[b]{.23\linewidth}
  \centering
  \includegraphics[width=\textwidth]{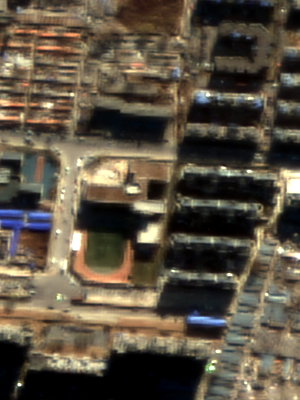}
  \centerline{(n) PSGAN \cite{PSGAN} }\medskip
\end{minipage}
\begin{minipage}[b]{.23\linewidth}
  \centering
  \includegraphics[width=\textwidth]{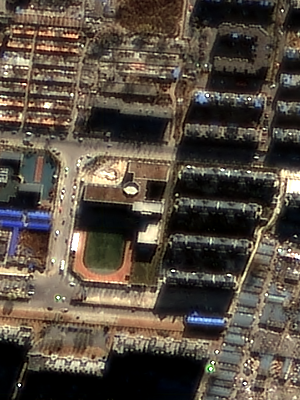}
  \centerline{(o) Pan-GAN \cite{Pan-GAN} }\medskip
\end{minipage}
\begin{minipage}[b]{.23\linewidth}
  \centering
  \includegraphics[width=\textwidth]{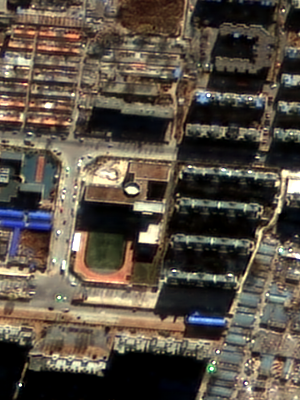}
  \centerline{(p) PGMAN }\medskip
\end{minipage}

\caption{Visual results on GaoFen-2 full-resolution dataset.}
\label{fig::GF-2 visual}
\end{figure*}

\begin{figure*}[t]
\centering

\begin{minipage}[b]{.23\linewidth}
  \centering
  \includegraphics[width=\textwidth]{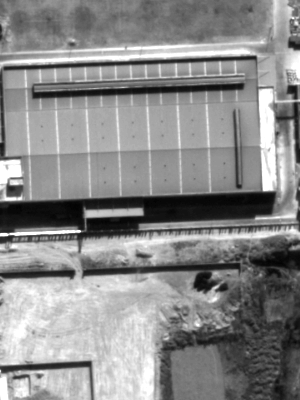}
  \centerline{(a) PAN}\medskip
\end{minipage}
\begin{minipage}[b]{.23\linewidth}
  \centering
  \includegraphics[width=\textwidth]{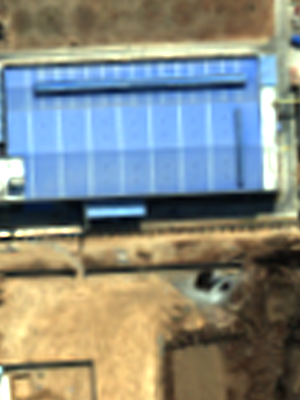}
  \centerline{(b) LR MS}\medskip
\end{minipage}
\begin{minipage}[b]{.23\linewidth}
  \centering
  \includegraphics[width=\textwidth]{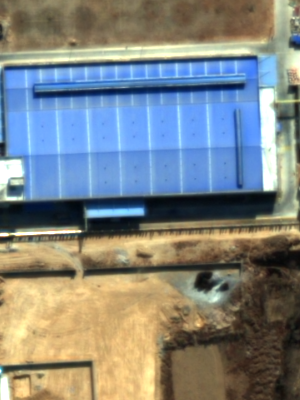}
  \centerline{(c) BDSD \cite{BDSD}}\medskip
\end{minipage}
\begin{minipage}[b]{.23\linewidth}
  \centering
  \includegraphics[width=\textwidth]{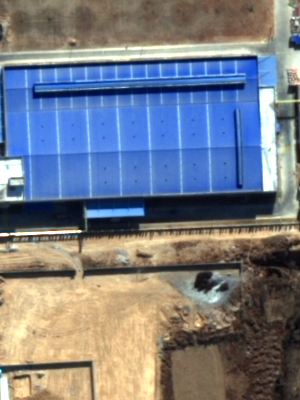}
  \centerline{(d) GS \cite{GS}}\medskip
\end{minipage}

\begin{minipage}[b]{.23\linewidth}
  \centering
  \includegraphics[width=\textwidth]{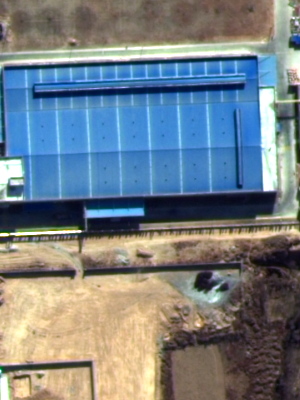}
  \centerline{(e) IHS \cite{IHS}}\medskip
\end{minipage}
\begin{minipage}[b]{.23\linewidth}
  \centering
  \includegraphics[width=\textwidth]{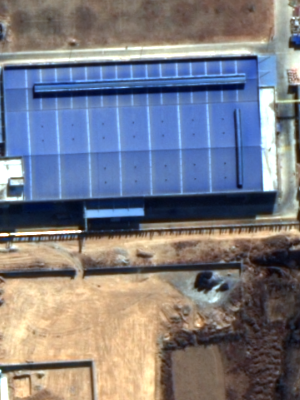}
  \centerline{(f) Brovey \cite{Brovey} }\medskip
\end{minipage}
\begin{minipage}[b]{.23\linewidth}
  \centering
  \includegraphics[width=\textwidth]{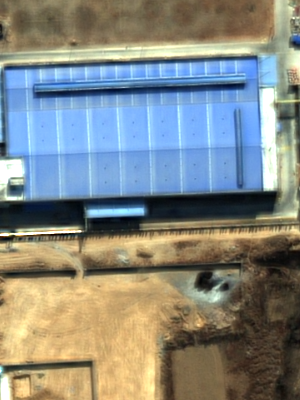}
  \centerline{(g) HPF \cite{HPFandHPFC} }\medskip
\end{minipage}
\begin{minipage}[b]{.23\linewidth}
  \centering
  \includegraphics[width=\textwidth]{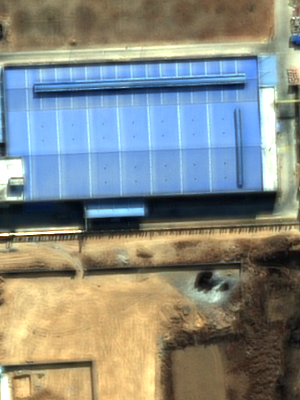}
  \centerline{(h) LMM \cite{LMVMandLMM} }\medskip
\end{minipage}

\begin{minipage}[b]{.23\linewidth}
  \centering
  \includegraphics[width=\textwidth]{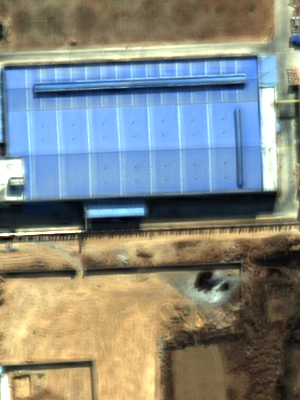}
  \centerline{(i) SFIM \cite{SFIM} }\medskip
\end{minipage}
\begin{minipage}[b]{.23\linewidth}
  \centering
  \includegraphics[width=\textwidth]{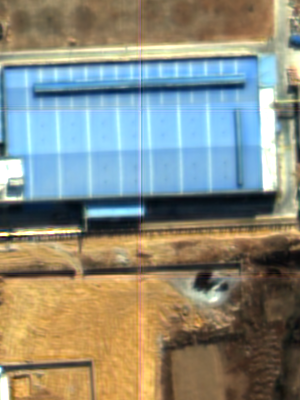}
  \centerline{(j) PNN \cite{PNN} }\medskip
\end{minipage}
\begin{minipage}[b]{.23\linewidth}
  \centering
  \includegraphics[width=\textwidth]{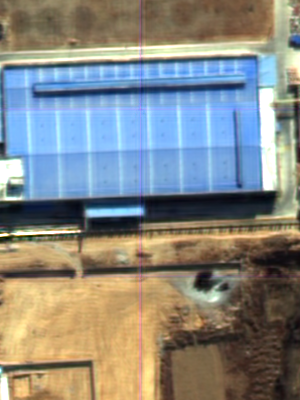}
  \centerline{(k) DRPNN \cite{DRPNN} }\medskip
\end{minipage}
\begin{minipage}[b]{.23\linewidth}
  \centering
  \includegraphics[width=\textwidth]{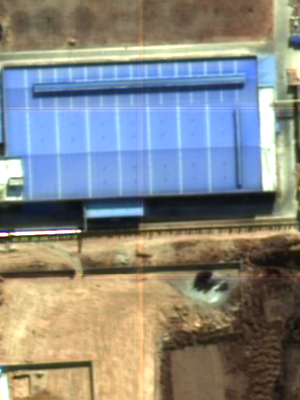}
  \centerline{(l) MSDCNN \cite{MSDCNN} }\medskip
\end{minipage}

\begin{minipage}[b]{.23\linewidth}
  \centering
  \includegraphics[width=\textwidth]{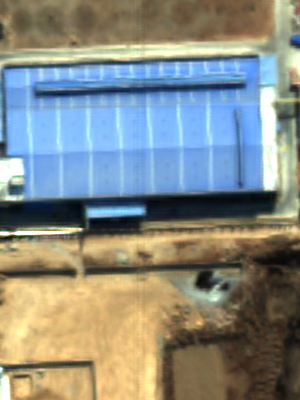}
  \centerline{(m) PanNet \cite{PanNet} }\medskip
\end{minipage}
\begin{minipage}[b]{.23\linewidth}
  \centering
  \includegraphics[width=\textwidth]{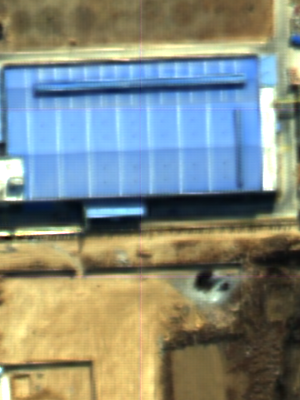}
  \centerline{(n) PSGAN \cite{PSGAN} }\medskip
\end{minipage}
\begin{minipage}[b]{.23\linewidth}
  \centering
  \includegraphics[width=\textwidth]{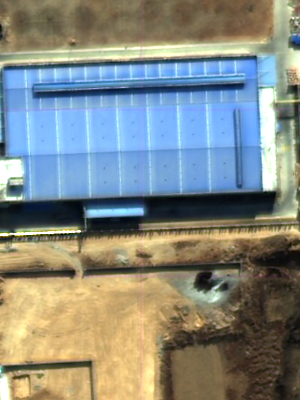}
  \centerline{(o) Pan-GAN \cite{Pan-GAN} }\medskip
\end{minipage}
\begin{minipage}[b]{.23\linewidth}
  \centering
  \includegraphics[width=\textwidth]{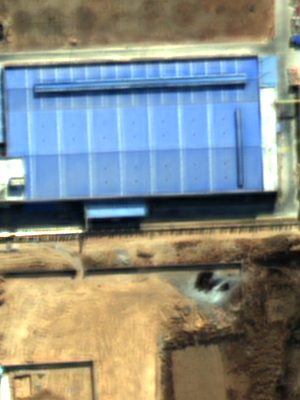}
  \centerline{(p) PGMAN }\medskip
\end{minipage}

\caption{Visual results on QuickBird full-resolution dataset.}
\label{fig::QB visual}
\end{figure*}

\subsection{Visual Results}

Fig.~\ref{fig::GF-2 visual} shows some example results on GaoFen-2 full-resolution images, which are produced from the original PAN and LR MS images. The results of Brovey (Fig.~\ref{fig::GF-2 visual}(f)), LMM (Fig.~\ref{fig::GF-2 visual}(h)) cannot preserve the spectral information from the LR MS image (Fig.~\ref{fig::GF-2 visual}(b)) very well. They tend to produce notable color distortions. Those supervised methods including PNN (Fig.~\ref{fig::GF-2 visual}(j)), DRPNN (Fig.~\ref{fig::GF-2 visual}(k)), MSDCNN (Fig.~\ref{fig::GF-2 visual}(l)), PanNet (Fig.~\ref{fig::GF-2 visual}(m)) and PSGAN (Fig.~\ref{fig::GF-2 visual}(n)) also demonstrate low quality spatial scenario details. It seems that the knowledge learned from the reduced-scale images is hard to generalize well to the real scenarios. This is because different from natural images, remote sensing images usually contain deeper bit depth and are with different pixel distribution. Down-sampling the original PAN and LR MS images in a supervised manner will change the data distribution and hurt the original information. However, when trained on the full-scale images, the quality is improved significantly for PGMAN (Fig.~\ref{fig::GF-2 visual}(p)). This gives us strong motivations to propose an unsupervised model since the pan-sharpening works are usually done on the full-resolution images in real applications.

Fig.~\ref{fig::QB visual} shows some example results on QuickBird full-resolution images. GS (Fig.~\ref{fig::QB visual}(d)), IHS (Fig.~\ref{fig::QB visual}(e)) and Brovey (Fig.~\ref{fig::QB visual}(f)) distort the spectral information from the LR MS image (Fig.~\ref{fig::QB visual}(b)) which can be seen from the color of the roof. The supervised methods including PNN (Fig.~\ref{fig::QB visual}(j)) and MSDCNN (Fig.~\ref{fig::QB visual}(l)) also produce unsatisfactory results with some distortions. Our method generates the pan-sharpened image (Fig.~\ref{fig::QB visual}(p)) with good spatial and spectral quality. The visual results from these two datasets demonstrate the superiority of our proposed model and show the advantage of unsupervised models for training on the original data distribution.

\subsection{Ablation Study}
We conduct ablation studies on GaoFen-2 dataset to verify the effectiveness of the proposed Q-loss. The quantitative results are listed in Table~\ref{tab::ablation results}. 

It is observed that using only the adversarial loss to optimize the network degrades the performance except for D$_\lambda$ index. Adversarial loss tries to simulate the generation of PAN and MS images, $i.e.$, the PAN and MS images are degradation version of the desired HR MS image along the spatial and spectral dimension, respectively. However, since the inverse of the process is ill-posed, even if the degraded images of the generated pan-sharpened images are identical to the corresponding PAN and LR MS, it is hard to guarantee that the pan-sharpened images are high fidelity in both spatial and spectral information. Q-loss makes an additional constraint to the learning, $i.e.$, the spatial and spectral information should be preserved across scales. Experimental results show that Q-loss produces better results than adversarial loss in most cases, indicating that Q-loss is a stronger constraint than adversarial loss. And finally, combining the two losses together further improves the performance, especially on the reference-based indicators. Therefore, with the help of Q-loss item and adversarial loss item, our proposed model PGMAN can be optimized to a good status and achieve competitive results on both reduced-scale and full-scale images.

Furthermore, we also conduct parameter analysis on GaoFen-2 Dataset, where we try 6 different combinations of $\alpha$ and $\beta$ values. Table~\ref{tab::parameter analysis} displays the results. At the same magnitude, the results of $\alpha=2\beta$ are better than those of $\beta=2\alpha$. It indicates that giving more weights to $D_1$ to make a balance is good for our model to learn, because the parameters of $D_1$ is less than $D_2$. When the magnitude is changed from $1e-3$ to $0$, we can find that both too small and too large values can affect the performance of the model. Considering all reference metrics and non-reference metrics, we finally choose $\alpha=2e-4$ and $\beta=1e-4$ to construct the loss function.

\begin{table}[t]\vspace{-10pt}
\centering
  \caption{Inference time and number of trainable parameters. Note that the pan-sharpened images are with a size of $400\times400\times4$, we give average time on them. As for the GAN-based models, we only take the parameters of the generator into account.}
  \label{tab::efficiency study}
  \begin{tabular}{r|r|c|c}
    \toprule
    Processor & Method & Time(s) & $\sharp$Params \\
    \midrule
    			& BDSD \cite{BDSD}		& 1.2613 & - \\
				& GS	 \cite{GS} 			& 0.1205 & - \\
    Intel Core	& IHS \cite{IHS} 		& 0.0121 & - \\
    i7-7700HQ	& Brovey \cite{Brovey}	& 0.0125 & - \\
    CPU@2.80GHz & HPF \cite{HPFandHPFC}	& 0.1061 & - \\
     			& LMM \cite{LMVMandLMM}	& 0.1062 & - \\
    			& SFIM \cite{SFIM}		& 0.1063 & - \\
    \midrule
    			& PNN \cite{PNN}			& 0.0001 & $\sim$ 0.080M \\
    			& DRPNN \cite{DRPNN}		& 0.0186 & $\sim$ 1.639M \\
    NVIDIA 		& MSDCNN \cite{MSDCNN}	& 0.0014 & $\sim$ 0.262M \\
    GeForce 	& PanNet \cite{PanNet} 	& 0.0004 & $\sim$ 0.077M \\
    RTX 2080Ti 	& PSGAN 	\cite{PSGAN} 	& 0.0008 & $\sim$ 1.654M \\
    			& Pan-GAN \cite{Pan-GAN}	& 0.0002 & $\sim$ 0.092M \\
				& PGMAN (ours) 			& 0.0004 & $\sim$ 0.385M \\
	\bottomrule
\end{tabular}
\end{table}

\subsection{Efficiency Study}
We evaluate the computational time and memory cost of ours and the comparison methods. All traditional methods are implemented using Matlab and run on an Intel Core i7-7700HQ CPU, and deep models are implemented in PyTorch and tested on a single NVIDIA GeForce RTX 2080Ti GPU. Table~\ref{tab::efficiency study} shows the inference speed and number of parameters of all comparative models. The inference speed is evaluated on 286 test samples, and each pair consists of a PAN image with $400 \times 400 \times 1$ pixels and a LR MS image with $100 \times 100 \times 4$ pixels. The time is calculated by averaging these 286 samples. Additionally, we report the number of trainable parameters of all deep learning related methods. 

BDSD is the slowest method, whose inference time is about 1.2613 seconds per image. IHS and Brovey only require about 0.01 seconds which are the fastest in traditional methods. 

Beneficial from the advance of GPU architectures, the inference time of deep learning models is satisfactory, almost all deep models only require less than 0.001 seconds to pan-sharpen one image, except for DRPNN model who is the largest deep model with deeper network architecture and much bigger convolution filters. PNN is the fastest since it consists of only three convolution layers. PGMAN is efficient in both model size and inference speed. 

\section{Conclusion} \label{section::Discussion}
In this paper, we propose a novel unsupervised generative multi-adversarial model for pan-sharpening, called PGMAN. PGMAN consists of one generator and two discriminators to reduce both the spectral and spatial distortions. To further improve the performance, we also introduce a QNR related loss to the unsupervised manner. As one of the advantages of unsupervised methods, our proposed method can be trained on either reduced-scale or full-scale images without ground truth. Our model is focused on original PAN and LR MS images without any preprocessing step to keep the consistency with the real application environment. The attractive performance on full-scale testing and satisfactory results on reduced-scale images demonstrate the powerful ability of our proposed model. 

However, there is still a gap between the supervised and unsupervised models on the reference metrics. When we use the GAN-based framework for spectral preservation, we simply down-sample the output to form a fake LR MS image for the discriminator to recognize. The method of down-sampling remains uncertain and may affect the performance of the model, which is also an ill-posed problem. There may be another better way to achieve the degradation process to discuss. In our future work, we will continue to study the architecture of unsupervised pan-sharpening models and further improve the performance.

\bibliographystyle{IEEEtran}
\bibliography{paper}

\end{document}